# Observations of the 2024 May 14 X8.7 Solar Flare with the Goldstone-Apple Valley Radio Telescope (GAVRT)

Thangasamy Velusamy[1], Ryan Dorcey[2], Nancy Kreuser-Jenkins[2], Lisa Nichole Lamb[2], Erica Pagano[2], Marin M. Anderson[1], Joseph Lazio[1], Steven Levin[1]

[1] Jet Propulsion Laboratory, California Institute of Technology, Pasadena, CA 91109, USA
[2] Lewis Center for Educational Research (LCER), Apple Valley, CA 92307, USA

## Abstract

The Goldstone-Apple Valley Radio Telescope (GAVRT) project conducts a regular monitoring program of the Sun. The GAVRT Solar Patrol project uses a 34 m diameter antenna to produce raster-scan maps of the Sun simultaneously at 4 frequencies ranging from approximately 3 GHz to 14 GHz. On 2024 May 14, as part of regular GAVRT Solar Patrol observations, raster maps were produced when an X8.7 solar flare occurred in active region AR13664. Here we present the GAVRT maps of the May 14 flare along with microwave flux density spectra showing the non-thermal microwave burst emission from mildly relativistic electrons produced in this largest flare of Solar Cycle 25 to date. AR13664 reappeared as AR13697 and continued to be very active, producing X flares while GAVRT monitored its activity. GAVRT microwave data provide a powerful complement to the energetic electrons tracked by X-ray, millimeter-wave and γ-ray emissions.

## 1  Introduction

The GAVRT (Goldstone-Apple Valley Radio Telescope) program is a partnership between JPL and the Lewis Center for Educational Research (LCER) that operates two retired NASA Deep Space Network (DSN) 34-m antennas at Goldstone, CA to provide science education and research opportunities to predominantly K-12 classrooms. Starting in 2021 a new science campaign dedicated to solar radio astronomy was formulated in order to expand the reach of GAVRT beyond the classroom to include the broader amateur science community (Velusamy et al. 2022). In the near term our aim is to use GAVRT Solar patrol to study and track solar activity during the interval leading up to and including solar cycle 25 (SC25) maximum while engaging amateur scientists in novel heliophysics research.

GAVRT Solar Patrol produces maps of the Sun, on a daily or higher cadence, at four frequencies between approximately 3 GHz and 14 GHz (centimeter wavelengths). During a flare, these frequencies track the gyro-synchrotron emission from electrons accelerated to energies of several hundreds of keV as they propagate in the magnetic fields above the active regions (e.g., Shaik & Gary 2021). Combined with X-ray observations that track energetic electrons with lower energies and millimeter-wave and γ-ray emission from higher energy electrons, the GAVRT Solar Patrol observations provide a means of tracing the energetics and magnetic field topology associated with active regions and the solar flares that they produce. On 2024 May 14, during regular GAVRT Solar Patrol observations, raster maps were being produced when an X8.7 solar flare occurred.

## 2 Observations and Results

GAVRT Solar Patrol observations are conducted with a receiving system that receives across the entire 3 GHz to 14 GHz band and produces four 100 MHz-wide sub-bands simultaneously (Velusamy et al. 2020, 2022). For the observations reported here, we used 3.5, 6.0, 8.45, and 13.0 GHz bands (Figure 1). For each time step during the raster scan, the antenna is pointed to a position offset relative to the center of the Sun. The antenna power for each band is recorded as counts. One complete solar map observation takes approximately 30 minutes. The maps shown in Figure 1 are constructed by gridding the time step data on to a regular $101 \times 101$ pixel grid (size $1.2° \times 1.2°$; pixel size $0.012°$) and applying triangulation interpolation. The map counts are then converted to brightness temperatures by subtracting any active region emission and calibrating the remaining quiet Sun counts to the quiet Sun brightness temperatures (Zirin et al. 1991). Figure 1 shows all three maps obtained on 2024 May 14 and one of the maps from May 06, when the Sun was relatively quiescent, for comparison.

The biggest flare (X8.7) in SC25 to date occurred on 2024-05-14 16:51 UT in the active region AR13664 (https://www.spaceweatherlive.com/en/archive/2024/05/14/xray.html), which was covered in the GAVRT maps observed on May 14 between 16:39 and 17:09 UT (Figure 1). The flare started around 16:46 UT (C8.8), peaked at 16:51 UT (X8.79), and lasted at least until 17:20 UT (M2). The GAVRT Solar Patrol raster scans covered the flare region AR13664 for about two minutes between 17:00 UT and 17:02 UT, close to the impulsive phase when the flare was at X5 and X4 levels, respectively. The flare region is very prominent, showing peak emission in the GAVRT maps at all four bands. In the last 17:17–17:47 UT map, the flare region was covered for two minutes from 17:38 to 17:40 UT, when the flare had almost ended. The GAVRT maps show only a marginal post-flare emission in this region. Unrelated to the X-flare, a long duration M4 (peak at 17:38 UT) flare occurred in the AR13680 region, which also is seen in the GAVRT map 17:17–17:47 UT (Figure 1). This flare region was covered in the raster scans from about 17:26 to 17:29 UT during its initial phase. Unlike the X-flare, it is marginally bright at the high frequency maps while brighter in low frequency bands.

We obtained flux densities at all four bands, integrating over an area containing AR13664 and AR13668, as marked by the box in Figure 1 (the large beam sizes at low frequencies includes both regions). The flux densities of the flares and active regions are determined by subtracting the quiet Sun emission (the residual after subtracting the emission from the bright flare/active regions from each map) in the region of integration. The right panels of Figure 1 show the resulting spectra. The flux density spectra are a useful diagnostic to quantify the microwave emission characteristics. In order to interpret the spectra of the flare and active region on May 14, we include an analysis of AR13664 on May 6, during a quiescent period. Comparing the quiescent maps on May 6 with the X-flare maps on May 14, we find:

(i) The cm-λ emission in the active regions is dominated by the gyro-resonance emission above sunspots (Kundu 1982). The May 6 quiescent active region spectrum resembles those from simulated gyro-resonance emission (Nindos et al. 1996), with flux densities decreasing with increasing frequency.

(ii) In contrast, the X-flare microwave spectra show a significant increase at high frequencies. The spectrum for the flare peak time also includes the active region emission; the flare contribution alone peaks above 10 GHz (as indicated by the down arrow in Figure 1), consistent with a microwave burst. The burst radiation during the impulsive phase is attributed to gyro-synchrotron radiation of the energetic (few hundred keV) electrons spiraling in the sunspot-associated magnetic field (Shaik & Gary 2021), typically peaking below or around 10 GHz, above which the spectral slope turns over from positive (the optically thick regime at lower frequencies) to negative (the optically thin regime at higher frequencies).

(iii) The pre- and post-flare spectra (Figure 1) depart from the quiescent thermal gyro-resonance spectrum with flux density excess at high frequencies, suggesting that the extremely high activity in AR13664 over this period has led to the buildup of a residual nonthermal component present in pre- and post-flare phases.

AR13664 has remained hyper-active and, since its return as AR13697 during the second rotation, has produced 88 C-, 27 M-, and 6 X-flares. GAVRT's Solar Patrol campaign will continue to monitor the Sun's activity at microwave wavelengths for the rest of SC25 and beyond.

**Acknowledgement:** This work was performed at the Jet Propulsion Laboratory (JPL), California Institute of Technology, under contract with the National Aeronautics and Space Administration.

## References

Kundu, M. R. (1965) *Solar Radio Astronomy,* (New York: Wiley Interscience)

Nindos, A., Alissandrakis, C. E., Gelfreikh, et al., 1996, *Solar Phys.*, 166, 55

Shaik, S. B. & Gary, D. E. (2021), *ApJ*, 919, 44J

Velusamy, T., Lamb, L., Dorcey, R., et al., 2022, AAS, iposter, https://aas240-aas.ipostersessions.com/?s=BB-D2-5D-AD-69-1F-AB-C4-DE-7D-AD-DE-8C-A6-BD-42

Velusamy, T., Kuiper, T. B. H., Levin, A. M., et al., 2020, *PASP*, 132: 094201

Zirin, H., Baumert, B. M. & Hurford, G. J. (1991), *ApJ*, 370, 779

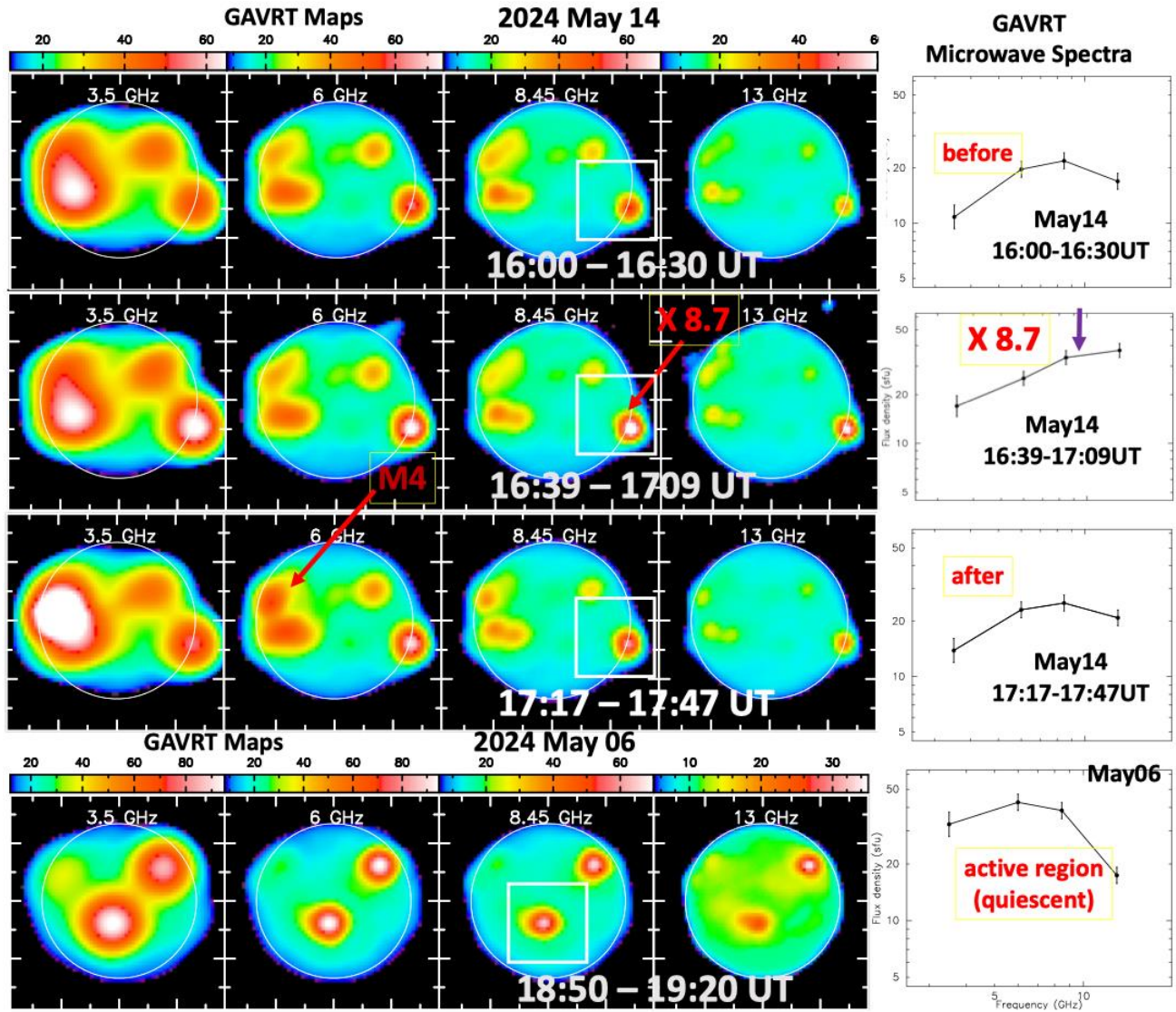

**Figure 1**. GAVRT Solar Patrol observations acquired during the 2024 May 14 X8.7 flare. The top three rows show GAVRT raster maps observed on May 14; the bottom row shows a comparable set of images acquired during a relatively quiescent interval on May 06. From left to right, the maps correspond to observing frequencies of 3.5, 6.0, 8.45 and 13 GHz. The arrows mark the X8.7 (16:51 UT; AR 13664) and M4 flares (17:38 UT; AR 13680). The white box represents the area containing active regions AR 13664 and 13668 used for integrated flux densities shown in the right-most panels. The microwave spectrum of the active region during a quiescent phase on May 06 is shown for comparison with the X flare on May 14. The variation in the spectra demonstrates the evolution from thermal gyro-resonance emission (May 06) to gyro-synchrotron emission from mildly relativistic electrons during the flare (May 14).

4